# In vitro reconstitution of microtubule dynamics and severing imaged by label-free interference reflection microscopy

Yin-wei Kuo[1,2], Jonathon Howard[2*]

1. Department of Chemistry
2. Department of Molecular Biophysics and Biochemistry, Yale University, New Haven, Connecticut, USA

*Correspondence: joe.howard@yale.edu

**Running head: Microtubule dynamics and severing using IRM**

**Abstract**

The dynamic architecture of the microtubule cytoskeleton is crucial for cell division, motility and morphogenesis. The dynamic properties of microtubules – growth, shrinkage, nucleation and severing - are regulated by an arsenal of microtubule-associated proteins (MAPs). The activities of many of these MAPs have been reconstituted in vitro using microscope assays. As an alternative to fluorescence microscopy, interference-reflection microscopy (IRM) has been introduced as an easy-to-use, wide-field imaging technique that allows label-free visualization of microtubules with high contrast and speed. IRM circumvents several problems associated with fluorescence microscopy including the high concentrations of tubulin required for fluorescent labeling, the potential perturbation of function caused by the fluorophores, and the risks of photodamage. IRM can be implemented on a standard epifluorescence microscope at low cost and can be combined with fluorescence techniques like total-internal-reflection-fluorescence (TIRF) microscopy. Here we describe the experimental procedure to image microtubule dynamics and severing using IRM, providing practical tips and guidelines to resolve possible experimental hurdles.

**1 Introduction**

Microtubules are cytoskeletal filaments that undergo stochastic transitions between growing and shrinkage phases, a feature termed dynamic instability. Dynamic instability can be described by four parameters: the growth and shrinkage rates, the catastrophe frequency (transition from growth to shrinkage) and the rescue frequency (transition from shrinkage to growth). The dynamics of microtubules is tightly controlled and is critical for many biological functions. Over the past few decades, numerous microtubule-associated proteins (MAPs) have been shown to modify microtubule dynamics in response to different cellular needs [1–6].

Total-internal-reflection-fluorescence microscopy (TIRF) assays have served as a crucial method for the discovery and functional characterization of MAPs, including polymerases [7], depolymerases [8–10], microtubule end tracking proteins [11] and motors [12–15]. However, fluorescence imaging has several drawbacks. First, it requires labeling by fluorophores, which can perturb microtubule properties or their interactions with MAPs [16, 17]. Second, photodamage can produce lattice defects, which can affect microtubule dynamics or even break the fluorescent microtubules [18, 19]. Photodamage can complicate the characterization of activities such as microtubule severing, and photobleaching limits the total light exposure to the specimen (shutter open time multiplied by the number of frames). And third, even though recent advances have provided new technologies to purify assembly-competent tubulin with origins other than mammalian brain tissues [20–29], it remains challenging to obtain sufficient amounts of starting materials to perform fluorophore conjugation.

Label-free imaging of microtubules using darkfield or differential interference contrast (DIC) microscopy has been the standard solution to circumvent the drawbacks posed by fluorescence microscopy [30–32]. However, the contrast variability of microtubules in DIC microscopy, and the high sensitivity of darkfield microscopy to stray light caused by minor misalignments and impurities in the solution or on the surface complicate their usage.

As an alternative, interference-reflection microscopy (IRM) provides a high-speed, easy-to-use, wide-field imaging method to visualize single microtubules with high contrast without the aforementioned limitations [33, 34]. In IRM, the image of a microtubule on the cover-glass surface is formed by the interference between the light reflected from the glass-solution interface and that from the solution-microtubule interface (Figure 1). IRM can be easily implemented at low cost by incorporating a 50/50 mirror to any epifluorescence microscope and requires only one-time alignment (Mahamdeh and Howard, 2019). In addition, a high numerical-aperture condenser is not needed in an IRM setup and the sample is thus more accessible to experimental manipulations. IRM can also be used in conjunction with TIRF microscopy to visualize both unlabeled microtubules and fluorophore-tagged MAPs as exemplified recently [35, 36]. In this chapter, we describe a protocol for imaging microtubule dynamics in vitro using IRM and its application to study microtubule severing.

**2 Materials**

2.1 Spastin purification

1. *Drosophila* His$_6$MBP-spastin short isoform (208aa to end) expressed in E. coli (Rosetta-DE3 competent cells 1 to 2 liters of culture) overnight at 16 °C. Cell pellets are stored at -80 °C. Maltose-binding protein (MBP) as a fusion protein improves the stability and solubility during expression and purification. A PreScission protease cleavage site is engineered between MBP and spastin to allow tag removal. A KKCK tag can also be introduced at the N-terminus of spastin to allow for downstream fluorescent labeling with maleimide-conjugated fluorophores. See [35] for detailed methods.
2. HisTrap™ HP column, 1mL.
3. MBPTrap™ HP column, 5 mL.
4. Superdex® 200 Increase 10/300 GL size exclusion column.
5. Fast protein liquid chromatography (FPLC) system. (e.g., Bio-rad NGC chromatography system).

6. His-binding buffer: 30 mM HEPES, 300 mM NaCl, 10 mM imidazole, 5% glycerol, 10 µM ATP, pH 7.4.

7. His-elution buffer (HisTrap): 30 mM HEPES, 0.3 M NaCl, 0.5 M imidazole, 10 µM ATP, 5% glycerol, 2 mM dithiothreitol (DTT) pH 7.4.

8. MBP-binding buffer: 30 mM HEPES, 300 mM NaCl, 5 mM $MgSO_4$, 10 µM ATP, 2 mM DTT pH 7.4.

9. MBP-elution buffer: 30 mM HEPES, 300 mM NaCl, 5 mM $MgSO_4$, 10 mM maltose, 10 µM ATP, 2 mM DTT pH 7.4.

10. Dialysis buffer: 30 mM HEPES, 5 mM $MgSO_4$, 300 mM KCl, 10% glycerol, 0.2% β-mercaptoethanol (BME), 10 µM ATP, pH 7.4.

11. SEC buffer: BRB80, 0.2 M KCl, 1 mM DTT, 10 µM ATP, pH 6.9. (*see* **Note 1**).

12. Protease inhibitors (stock solution): 200 mM Pefabloc in $ddH_2O$, and 5 mg/mL leupeptin in DMSO. Aliquots stored at -20°C. Alternatively, other commercially available EDTA-free protease inhibitors mix can be used.

13. Benzonase (can be replaced by DNase I).

14. 10 mg/mL lysozyme in phosphate buffer saline, 500 µL aliquots stored at -20 °C (stock solution).

15. PreScission protease.

16. β-mercaptoethanol (BME).

17. 25 mM Adenosine triphosphate (ATP) disodium salt in $ddH_2O$, adjusted pH to ~7. Aliquots of 5 µL stored in -20 °C (stock solution).

18. Amicon ultra centrifugal filter, 30 kDa cut-off.

2.2 Chamber preparation

1. Piranha-cleaned and silanized #1.5 cover glasses (one 18 mm × 18 mm and one 22 mm × 22 mm). For cover glass cleaning and silanization see [37].
2. Parafilm or double-sticky tape.
3. Antibeta-tubulin SAP.4G5 (Sigma-Aldrich, T7816).
4. 1 mg/mL anti-tetramethylrhodamine (Invitrogen, Molecular Probes, A6397). Stored as 5 µL aliquots at -20 °C.
5. 1 mg/mL anti-biotin antibody (Sigma-Aldrich, B3640). Stored as 5 µL aliquots at -20 °C.
6. BRB80 buffer: 80 mM piperazine-N,N'-bis(2-ethanesulfonic acid) (PIPES)-KOH, pH 6.9, 1 mM ethylene glycol-bis(2-aminoethylether)-N,N,N',N'-tetraacetic acid (EGTA), 1mM $MgCl_2$. Use KOH rather than NaOH to adjust pH as sodium ions increase the hydrolysis rate of GMPCPP in the microtubules [38]. Buffer is filtered (0.22 µm), degassed and stored at 4 °C.
7. Casein. Add 20 mL BRB80 to ~0.5 g casein powder in a falcon tube and gently shake to dissolve at 4 °C overnight. Afterwards, centrifuge at 4,000 *g* at 4 °C for 10 min. Filter the supernatant (0.22 µm) and adjust to 10 mg/mL with cold BRB80. Stock solution stored at -20 °C as 10 µL aliquots.
8. F127. 1% pluronic F127 (w/v) is dissolved in BRB80 overnight, filtered (0.22-µm syringe filter), and stored at -20°C in 100 µL aliquots.

2.3 Dynamic and severing assays

1. Interference-reflection microscopy is set up on an inverted TIRF microscope with 100X high numerical aperture objective [33]. IRM can be implemented to an epifluorescence microscope by inserting a 50/50 mirror in the light path (Figure 1). For detailed procedure of IRM implementation and alignment, see [39].

2. Temperature-control system such as objective heater or microscope chamber with temperature-feedback.

3. Tubulin, unlabeled (porcine or bovine brain). Purified mammalian brain tubulin is commercially available (e.g. Cytoskeleton Inc.) (*see* **Note 2**).

4. Biotinylated tubulin (Cytoskeleton Inc., Denver, CO, USA, T333-B). See [40] for a labeling procedure.

5. Tubulin, carboxytetramethylrhodamine (TAMRA)-labeled (we label tubulin purified from bovine brain as described in [40]). Rhodamine-labeled porcine tubulin is also available commercially (Cytoskeleton Inc.).

6. 10 mM Guanosine-5'-[(α,β)-methyleno]triphosphate (GMPCPP) in ddH$_2$O stored as 5 µL aliquots at -20°C (stock solution).

7. 100 mM of MgCl$_2$ in ddH$_2$O. Filtered by 0.22 µm syringe filter (stock solution).

8. 1 mg/mL (4.2 µM) catalase in BRB80, aliquots, snap-frozen in liquid nitrogen, stored at -80°C (stock solution).

9. 500 mM Dithiothreitol (DTT). Aliquots stored at -20°C (stock solution).

10. 2 M D-Glucose in water, 10 µL aliquots, stored at -20°C (stock solution).

11. 2 mg/mL (12 µM) Glucose oxidase in BRB80, 10 µL aliquots, snap-frozen in liquid nitrogen, stored at -80°C (stock solution).

12. 25 mM Guanosine triphosphate in ddH2O, adjusted pH to ~7. Aliquots of 5 µL stored in -20 °C (stock solution).

13. Tetraspeck fluorescent microspheres (0.2 mm, Invitrogen, #T7280): 1:100 dilution in BRB80.

14. Ultrapure Water (resistivity > 18MΩ·cm).

**3 Methods**

3.1 Purification of recombinant spastin

(adapted from [35] with modifications)

1. Resuspend the cell pellet in cold His-binding buffer (30 mL buffer/ liter of culture) with addition of protease inhibitors (final concentration 0.2 mM of Pefabloc and 5 µg/mL leupeptin), DTT (final concentration 2 mM) and lysozyme solution (final concentration 0.1 mg/mL). All steps of the purification should be performed at 4 °C in a cold room.

2. Sonicate the cell suspension on ice or break the cells with a homogenizer.

3. Add benzonase (0.3 U/µL final concentration) and incubate the solution on ice for 10 min. Transfer into the centrifuge tube and spin down at 20,000 *g* for 30 min at 4 °C. Collect the supernatant that contains soluble protein fraction.

4. Wash the HisTrap column (1 mL) with 5 column volumes (CV) of water and equilibrate with 10 CV of His-binding buffer. Load the supernatant onto the HisTrap column.

5. Wash the column with His-binding buffer + 10% His-elution buffer for 10 CV. Elute the protein with elution buffer gradient (10-75% over 30 CV) and check SDS-PAGE for purity. Check the eluted fractions with SDS-PAGE and pool the peak fractions (*see* **Note 3**).

6. See Note in step 5. Dilute the fractions with MBPTrap binding buffer to < 50 mM imidazole, and load onto the MBPTrap (5 mL) column (flow rate 0.5 to 1 mL/min).

7. Wash the column with MBP-binding buffer for 10 CV and elute with 5 CV elution buffer.

8. Measure the protein concentration by the absorbance at 280 nm and dilute with dialysis buffer to a final concentration of less than 1 mg/mL. Add PreScission protease (1 µL protease/100 µg of protein) to the protein solution and dialyze overnight at 4°C to allow the tag cleavage and buffer exchange (*see* **Note 4**).

9. Verify the tag cleavage with SDS-PAGE. If the tag is removed, filter the solution using a low-protein-binding syringe filter to remove possible aggregates.

10. Equilibrate the 1 mL HisTrap column with His-binding buffer. Dilute the overnight-dialyzed sample with 1 volume of His-binding buffer (no DTT) to decrease the BME concentration to 0.1% and filter using a low-protein-binding syringe filter.

11. Flow the solution through the HisTrap column to remove the cleaved His$_6$MBP-tag and collect the flow through. Wash the HisTrap column with 3 to 5 CV of His-binding buffer + 7.5% His-elution buffer and combine with the flow-through.

12. Concentrate the collected solution for size-exclusion chromatography with a 30-kDa Amicon filter. Inject into the size exclusion column (pre-equilibrated with SEC buffer) and elute with SEC buffer.

13. Check the purity with SDS-PAGE and combine fractions with high purity. Measure the protein concentration and concentrate it with an Amicon if needed.

14. Spin the solution with an ultracentrifuge at 20,000 *g* for 10 min at 4°C to remove potential aggregates.

15. Collect the supernatant and measure the concentration by absorbance. Dilute with SEC buffer if needed. Store the solution at 5 to 10 µM.

16. Make 5 µL aliquots, flash-freeze in liquid nitrogen and store at -80°C.

3.2 Microtubule dynamic assay with IRM

Figure 1 is a schematic of the IRM microtubule dynamic assay setup.

3.2.1 Preparation of GMPCPP-stabilized microtubules

GMPCPP-stabilized microtubules are frequently used as seeds for growing the dynamic microtubule extensions. They can also be used as substrates in vitro severing assays. The average length of GMPCPP-microtubules can be tuned by controlling the polymerization time.

1. In a 0.6 mL centrifuge tube, mix 10 µL of 10 µM TAMRA-labeled tubulin (20-30% labeling stoichiometry determined by the ratio of 280 nm absorbance to the wavelength of dye's maximal absorption, see **Note 5**), 5 µL of 10 mM GMPCPP solution, 0.5 µL 100 mM $MgCl_2$, and 34.5 µL BRB80 to give a 2 µM final tubulin concentration.

2. Incubate the mix on ice for 5 min and transfer to 37 °C for 25 min to polymerize the microtubules. The microtubules are typically around 3-4 µm in length, which is suitable for seeds. For severing assays where longer microtubules are desired, incubate at 37 °C for 1-2 hours.

3. Dilute the mix with 100-350 µL of room temperature BRB80 to stop the polymerization. Spin down the solution at room temperature using an ultracentrifuge (126,000 g, 5 min) to remove unpolymerized tubulin. Discard the supernatant and resuspend the pellet gently in 200 µL room temperature BRB80 using a pipette whose tip has been cut off to reduce filament breakage due to the shear forces (see **Note 6**). GMPCPP-microtubules can be stored for a few hours at room temperature. The typical depolymerization rate at room temperature of these GMPCPP-microtubules is about 0.02 µm/min; stability can be further enhanced by cycling them for the second time in the presence of GMPCPP as described in [37].

3.2.2 Assembly of the flow chamber

A simple flow chamber can be constructed by adhering an 18×18 mm cover glass to a 22×22 mm cover glass using strips of parafilm or double sticky tape in between. A small metal holder is designed to fit the microscope stage as described in [37], allowing the objective to approach the 22x22 mm cover glass from underneath for imaging.

1. Cut several strips of parafilm (~30 mm long and 2 mm wide) with a razor blade.
2. Place the silanized 22×22 mm cover glass on a clean lens paper and arrange the parafilm strips on the top side of the cover glass to form a series of parallel channels that are about 3 mm wide each. Center the channels as much as possible. Cut off the parafilm strips that protrude beyond the cover glass with a razor blade. Typically, two to three channels can be made on a single cover glass. Each channel volume is normally 5 to 6 µL (*see* **Note 7**).
3. Place the silanized 18×18 mm cover glass on top of the parafilm channels and gently apply pressure on the 18×18 mm cover glass at the parafilm regions with a clean plastic or wood tweezer to facilitate the adhesion. The 18×18 mm cover glass should be placed so that there is a 1 to 2 mm space on all sides of the 22×22 mm cover glass for clamping into the holder and exchanging the solution.
4. Place the assembled flow channel onto a 90 °C hot plate heated with a clean lens paper beneath the 22×22 mm cover glass. Heat briefly until the parafilm strips become translucent. Alternatively, a heated brass density cube can be placed on to the 18×18 mm side briefly to melt the parafilm (*see* **Note 8**).

5. Mount the assembled channel onto the metal holder. Take extra care to ensure the channel lays flat on the holder without tilt (*see* **Note 9**).

3.2.3 Microtubule binding and surface passivation

Stabilized microtubules are bound to the cover glass surface with a spacer protein that is adsorbed to the surface via non-specific hydrophobic interactions. The selection of the spacer protein depends on the specific design of each experiment. Typically, we use anti-TAMRA antibody to bind TAMRA-labeled GMPCPP microtubules onto the surface in the dynamic assay. This has the advantage in an IRM dynamic assay since the fluorophore labeling allows the distinction of GMPCPP-microtubule segments using TIRF from the unlabeled microtubule extension. We found that 20% to 30% of TAMRA labeling stoichiometry is ideal to affix the GMPCPP-microtubules seeds onto the antibody coated surface firmly. Alternatively, biotinylated GMPCPP-microtubules with 10% to 15% labeling stoichiometry (*see* **Note 10**) can be attached to a neutravidin- or anti-biotin antibody-coated surface if the seeds do not need to be marked fluorescently. The location of GMPCPP-microtubule seeds can also be identified by taking a snapshot of the field of view before flowing in the polymerization reaction mix.

1. Perfuse 40 µL of BRB80 at room-temperature BRB80 (to avoid air bubbles) using a vacuum to draw the solution into each flow cell. Switch to filter paper after the flow channel is filled with solution for subsequent perfusion steps.
2. Perfuse 20 µL of antibody solution through each channel. We typically use 1:50 dilution in BRB80 for the anti-TAMRA antibody. Leave a drop of solution at both sides of each channel to prevent it from drying. Incubate the solution in the channel for 10 min inside a humidity chamber at room temperature (*see* **Note 11**).

3. Perfuse 40 µL of room temperature BRB80 through each channel to wash off the unbound antibody.

4. Perfuse in 20 µL of a 1% F127 solution and incubate for 30 min to block the surface. Wash the flow channel by perfusing though 40 µL of room-temperature BRB80. Pluronic F127 is a triblock copolymer that consists of a central polypropylene oxide region, which adsorbs strongly to the hydrophobic silanized surface, and two flanking polyethylene glycol extensions, which form a layer of "PEG brushes" that prevent non-specific protein adsorption onto the cover-glass surface.

5. Perfuse in 20 µL of 2 mg/mL casein solution (in BRB80) and incubate for 10 min at room temperature to provide a secondary passivation of the surface. Wash the channel again with 40 µL of room temperature BRB80 (*see* **Note 12**).

6. Surface-passivated flow channels can be stored in the humidified chamber at room temperature for one to two hours before use.

3.2.4 Imaging microtubule dynamics

1. Set the objective heater (or other temperature control system) to the desired temperature, depending on the protein of interest and the source of tubulin. For mammalian brain tubulin, 35 °C is typically used for imaging dynamic microtubules. Polymerization of brain tubulin at 28 °C is often assayed when using MAPs from yeast or *Drosophila*. The actual temperature in the flow chamber can be measured using a thin thermal probe.

2. Choosing the imaging condition based on the parameters to be measured. For growth rates and catastrophe frequencies, 0.2 frames per second (fps) with a total of 20 min of imaging suffices. Higher temporal resolution of 5 to 10 fps is necessary to visualize rapid microtubule shrinkage. Depending on the light source and the camera used, temporal resolution of 100 to 1000 fps (or higher) can be feasible. If the sample is imaged by both TIRF and IRM, the frame rate will be limited by the rate of switching of the filter cubes (typically about 1 second for each switching step). An optimal light intensity for IRM imaging nearly saturates the dynamic range of camera when the exposure time is 10 ms. Averaging of images will reduce background noise and improve the image quality [33]. We installed a long pass filter (> 600 nm) in the illumination light path to minimize the risk of photodamage caused by visible light (if any fluorophore is used) and UV light.

3. Place the flow chamber onto the microscope stage. Dilute the GMPCPP-microtubules solution 10 times with BRB80 (prewarmed to imaging temperature). Flow in 20 µL of diluted microtubule solution and watch the landing of microtubules in real time with IRM imaging (*see* **Note 13**).

4. Wash the flow channel with 40 µL of BRB80 (prewarmed to the imaging temperature).

5. Acquired a background image (for background subtraction) by taking 100 frames without delay (i.e., streaming near 100 fps with 10 ms exposure time) while moving the microscope stage with a stage controller or computer software. The median of the 100 images will be used as a background image (see section 3.3 for details). The sample should lie flat without tilt on the microscope stage to prevent erroneous background due to variation in axial positions (*see* **Note 14**).

6. Quickly warm (to the imaging temperature) the polymerization mix that contains 7-15 µM unlabeled tubulin (freshly thawed), 1 mM GTP and 1 mM DTT and the protein-of-interest (if applicable) in BRB80. Flow in 20 µL of polymerization mix into the flow cell and start image acquisition. Take special care to prevent the flow cell from drying during the imaging (*see* **Note 15**). Microtubule growth from the GMPCPP-seeds is typically observed at a tubulin concentration above ~7µM. Spontaneous nucleation of microtubules can occur above ~15 µM and may interfere with imaging (*see* **Note 16**). The ionic strength of the reaction buffer can be adjusted by addition of KCl (alternatively BRB20 buffer, where 20 mM PIPES is used, can be used to lower the ionic strength) based on the MAPs studied. If TIRF imaging is used, include oxygen scavengers in the polymerization mix (we typically use 40 mM glucose, 40 mg/mL glucose oxidase, 16 mg/mL catalase, 0.1 mg/mL casein, 10 mM DTT). When supplementing the reaction with oxygen-scavenger mix, add glucose oxidase last as it initiates the oxidation reaction of glucose. A 2x oxygen scavenger mix can be prepared in advance and kept on ice for ~30 min. For optimal performance, prepare fresh oxygen scavenger mix before each reaction. Figure 2 shows an example of combining TIRF and IRM to visualize fluorescently labeled MAPs (EB1-GFP) and unlabeled dynamic microtubules (*see* **Note 17**).

3.3 Microtubule severing assay with IRM

In this section, we describe the *in vitro* microtubule severing assay using unlabeled stabilized microtubules affixed on the surface of a microscope coverslip. This eliminates the photodamage effects in a TIRF microtubule severing assay. Photodamage can either induce breakage of microtubule filament directly or produce lattice defects that makes the filament more susceptible to severing, thus complicating the study of severing enzyme activities. Here we described the protocol using GMPCPP-stabilized microtubules as substrates. Similar experimental procedure can also be applied to microtubules stabilized by taxol, or GMPCPP-tubulin capped GDP-microtubules.

1. Prepare the flow chamber as described in sections 3.2.2 and 3.2.3. Use anti-tubulin antibodies (50 times dilution in BRB80) instead of anti-TAMRA antibodies to bind unlabeled microtubules (*see* **Note 18**).
2. Set the objective heater to the desired temperature. For *Drosophila* spastin, set the temperature to 28°C.
3. Flow unlabeled GMPCPP microtubules into the flow channel using a cut pipette tip (*see* **Note 6**) and monitor microtubule density on the surface by IRM. Wash with 40 µL room temperature BRB80 when desired density is reached.

4. Prepare a severing-enzyme reaction mix. For *Drosophila* spastin, we use the following conditions: 1 mM MgATP, 5 mM DTT, 50 mM KCl, 1 to 200 nM spastin in BRB80 (*see* **Note 19**).

5. Incubate the severing reaction mix at 28 °C for 1 min. Perfuse 20 µL of severing reaction mix into the flow cell and start image acquisition. An example of GMPCPP-stabilized microtubules severed by spastin is shown in Figure 3. The selection of frame rate will depend on the concentration of the severing enzyme in the reaction. For concentrations < 10 nM, 0.5 to 1 fps is normally sufficient to capture the severing events (*see* **Note 20**).

3.3 Image processing and data analysis

3.3.1 Quantification of microtubule dynamics

1. Using image analysis software such as ImageJ/Fiji, generate a background image by taking the median projection of the 100-frame image stack taken in section 3.2.4 step 3. The background image retains the illumination profile and stationary features (e.g. dirt on the camera or in the light path) but excludes the nonstationary features such as microtubules on the surface.

2. Subtract the background image from each frame of the movie (*see* **Note 21**).

3. For measurements of microtubule dynamic parameters, generate a kymograph for each microtubule. A kymograph is a time-stacked image of an individual microtubule filament and can be produced by using the "Reslice" tool or kymograph plug-ins in Fiji/ImageJ (*see* **Note 22**).

4. Measure the parameters of microtubule dynamics from the kymographs. The growth rate and the shrinkage rate are measured as the slopes of the growth and shrinkage phases. Microtubule polarity can usually be determined by comparing the growth rates at the ends (the plus end grows faster) (*see* **Note 23**). Catastrophe frequency is normally calculated as the total number of catastrophe events observed divided by the total growth time from all microtubules analyzed. Similarly, rescue frequency can be determined by the number of rescue events divided by the total shrinkage time. Alternatively, catastrophe frequency can be determined by the lifetime from the onsets of microtubule growth from the seeds until a catastrophe event occurs [10, 41]. Note that the maximum lifetime is restricted by the length of the experiment. Several initial trials should be performed to determine the suitable image acquisition time to avoid underestimating the catastrophe frequency.

3.3.2 Quantification of microtubule severing

1. Perform background subtraction as described in section 3.3.1.
2. The severing rate can be measured in two ways. One common way is to count the number of microtubule breakage events and divide this value by the total microtubule length and total time analyzed [42]. One drawback of this approach is that the severing events are not uniformly distributed over time and there is a lag time before the severing occurs [43–45]. Another disadvantage is that the short microtubule fragments generated by severing have a higher chance to detach from the surface over time and cannot be visualized. This makes the measurement at higher concentration of severing enzyme difficult due to the fast severing and the detachment of microtubule fragments. Alternatively, severing activity can be measured as the pre-severing time normalized by the microtubule length (the time from perfusing in severing reaction mix until the first severing event occurs on each microtubule divided by the microtubule length).

# 4 Notes

1. If fluorescent labeling with maleimide-conjugated fluorophores is desired, substitute DTT with 0.5 mM tris(2-carboxyethyl)phosphine (TCEP).
2. The tubulin should undergo a polymerization and depolymerization cycle to remove denatured tubulin and tubulin aggregates before making 40 µM aliquots, flash-freezing them and storing at -80 °C. Cycling is especially important if lyophilized tubulin is used.
3. If the purity of the eluted fractions is good, omit the MBPTrap affinity chromatography step and proceed to step 8 directly. The MBP-column is mainly used to improve the purity if the His-eluted fractions contain significant amounts of non-specific binding proteins or degradation products.
4. We found that MBP-cleaved spastin has lower solubility: therefore, it is crucial to maintain a high salt concentration and a low protein concentration to minimize protein precipitation after tag cleavage.

5. The tubulin concentration and labeling density can be determined from the absorbance at 280 nm ($A_{280}$) and at the absorption maximum of the fluorophore ($A_\text{fluor}$) using the following formulas: $[\text{tubulin}] = (A_{280} - CF \times A_\text{fluor})/\varepsilon_\text{tubulin}$, $labeling\ density = \frac{A_\text{fluor}}{\varepsilon_\text{fluor}[tubulin]}$, where $\varepsilon_\text{tubulin}$ is the molar extinction coefficient of tubulin at 280 nm, $\varepsilon_{fluor}$ is the molar extinction coefficient of the fluorophore at the absorption maximum. CF is the correction factor of the fluorophore, which is the ratio of the fluorophore absorbance at 280 nm and at the maximum. The tubulin is typically labeled with a density of 40-80 % and diluted with unlabeled tubulin to 25-30% final labeling density.

6. Mechanical damage caused by shear force during pipetting can affect the apparent severing rate. Use a cut tip and avoid vigorous pipetting when handling microtubules to reduce the effect of shear flow.

7. The parafilm strips should not be moved once they are in contact with the cover glass to keep the surface clean.

8. Avoid prolonged heat that can decrease surface hydrophobicity and may lead to cracking of the cover glass.

9. If a suitable cover glass holder is not available, a piranha cleaned and silanized microscope slide can be used to substitute the 22×22 mm cover glass and the flow channel can be constructed in a similar manner. However, this will require the inversion of the assembled channel on an inverted microscope and is thus not suitable for exchanging solution directly on the microscope.

10. Biotin labeling density can be determined by using commercial biotin quantification kit (e.g., Pierce) with either fluorescent or colorimetric detection method.

11. A simple humidity chamber can be made by using an empty pipet tip box filled with a small amount of buffer. Using water will lead to dilution of the buffer, though the mixing inside the flow cell by diffusion will be very slow.

12. Passivation by F127 is only applicable to hydrophobic surfaces while casein can adsorb to both hydrophobic and hydrophilic surfaces. A double-blocking strategy using F127 and casein can improve non-specific protein binding and therefore reduce fluorescence background in TIRF. Addition of casein is particularly useful when higher concentrations of antibody or other spacer protein is used to promote stronger microtubule binding.

13. A well aligned IRM microscope should allow direct visualization of microtubules without background subtraction. A solution with 10 times dilution of GMPCPP-microtubules normally reach a suitable surface density within 1 to 2 min. A simple squash of the prepared microtubules solution can also be used to estimate the density of microtubules in the solution in advance. If the density on the surface is low after incubation of 5 min, another 20 µL of microtubule solution can be perfused in.

14. Alternatively, an averaged background image can be acquired before flowing in the microtubule seeds. This method requires the cover-glass surface to be as clean as possible with minimal drift during the video acquisition since any particles on the surface will be retained in the background image and can cause local artifact after background subtraction.

15. A simple way is to leave a drop of polymerization mix solution at each side of the channel, surround the flow channel holder with tissues wetted with buffer and put a lid on top of the microscope stage. This is normally sufficient for 15 to 20 min imaging. Alternatively, flow cells can be sealed using VALAP (a mixture of Vaseline, lanolin and parafilm with equal ratio) or vacuum grease, but sealing the channel prevents further solution exchange of the channel.

16. Tubulin loses activity over time and consistent handling of tubulin, such as the thawing time and incubation temperature is important for reproducible measurements.

17. Thermal fluctuations can cause longer microtubules to flicker away from the surface and make the tips difficult to see with IRM because the contrast switches between dark and white when the height changes by ~150 nm (wavelength/4). This is not a problem in the example shown in Figure 2 because the microtubules are relatively short. If the problem does occur, addition of a crowding agent such as 0.05% to 0.1 % methyl cellulose (4000 cP) reduces the thermal fluctuations. However, because crowding agents and solution viscosity can affect microtubule dynamics [46], careful control experiments should be performed.

18. The density of the antibody affects how strongly a microtubule is attached to the surface and can influence the severing rate. The concentration of antibody used should be the same for the control and assay conditions to minimize discrepancies resulting from the microtubule surface binding.

19. Full length spastin loses activity over time and should be used within 30 min of thawing. For reproducible measurements, freshly thawed spastin should be used for each reaction. If aggregation of the severing enzyme is observed, spin down the solution at 20,000 $g$ for 5 min at 4°C to remove the aggregates. Recheck the protein concentration by absorbance at 280 nm.

20. The severing rate increases steeply with spastin concentration. To measure the short pre-severing time when high concentration of severing enzyme is used, use a simple microfluidic system to control solution flow while imaging simultaneously to replace manual flow to avoid the delay between flowing in the severing reaction mix and the start of image acquisition.

21. If using Fiji/ImageJ, use the 32-bit (float) option for image subtraction as the objects are darker than the background and will appear as negative values after subtraction. Images can be converted to 16-bit or 8-bit if needed after the background subtraction.

22. Sample drift during imaging can hamper the accurate measurement of microtubule dynamics from the kymograph. Thermal drift can be reduced by mounting the arc lamp off the microscope and pre-equilibrating the sample holder and solutions at the imaging temperature for several minutes. Air flow (e.g. from the air-conditioner of the room) can also contribute to the sample drift, which can be reduced by providing the microscope with a barrier (e.g. a transparent plastic box). Drift correction can also be done by picking a static object on the surface such as a dirt particle as a landmark. For IRM/TIRF imaging, use Tetraspeck beads to align both TIRF and IRM channel after drift correction.

23. While the plus end growth rate is typically faster than the minus end growth rate, it is prudent to verify the microtubule polarity if the tubulin or the MAP used in the dynamic assay has not been characterized with respect to the polarity of growth. Microtubule polarity can be determined using the polarity-marked microtubule seeds [17, 47] or using a plus end-directed motor (e.g., kinesin-1) in a motility assay [15].

**Acknowledgements**

We thank Dr. Mohammed Mahamdeh and Dr. Anna Luchniak for comments and discussions on the manuscript. This work was supported by NIH grants R01 GM139337, DP1 MH110065, and R01 NS118884 (to J.H.) and a fellowship from the Ministry of Education in Taiwan (to Y-W.K.).

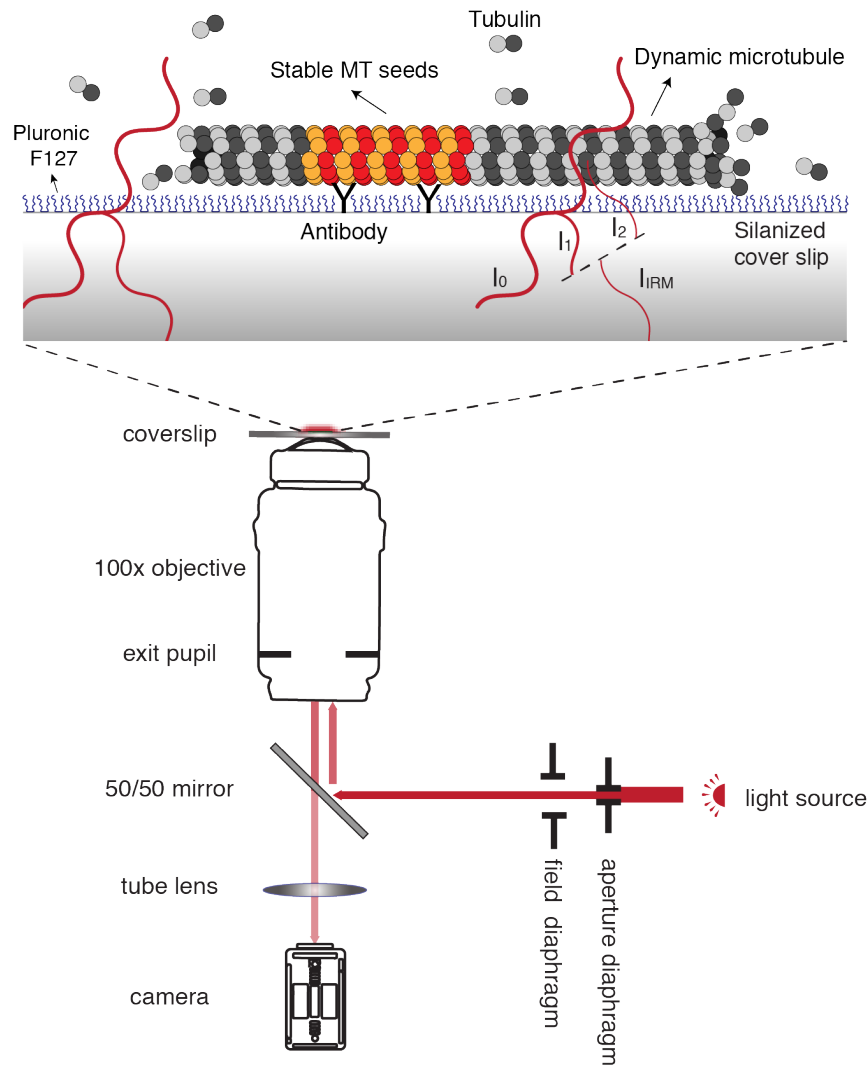

**Figure 1. Setup of the microtubule dynamic assay using interference-reflection microscopy (IRM).** GMPCPP-stabilized microtubule seeds (TAMRA-labeled) are affixed onto the surface of the coverslip by anti-TAMRA antibody. Dynamic microtubule extensions polymerized from unlabeled tubulin elongate from the microtubule seeds and are visualized by IRM. Epi-illumination from the light source is partially reflected by the 50/50 mirror to illuminate the sample. The interference between the light reflected from the water-microtubule interface and the water-glass interface provides the contrast to visualize microtubule filaments over background. Figure modified from [39].

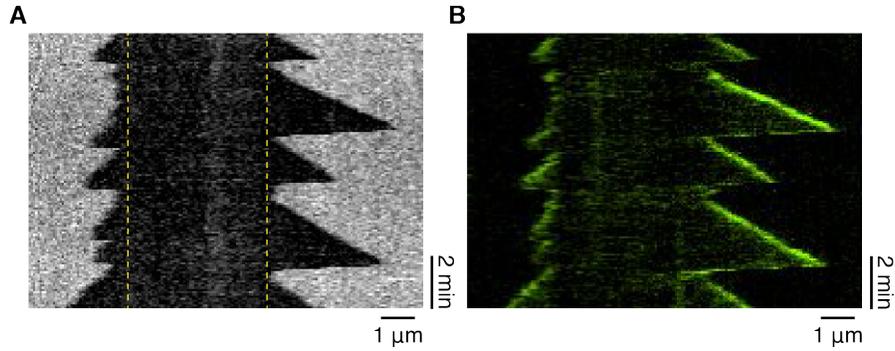

**Figure 2. Unlabeled dynamic microtubules and fluorescently labeled MAPs imaged by IRM and TIRF microscopy.** (A) Kymograph of a dynamic microtubule (unlabeled) in the presence of EB1-GFP in IRM channel. The faster-growing right end of the microtubule is the plus end. Yellow dashed lines indicate the position of the GMPCPP-stabilized microtubule seed (TAMRA-labeled). (B) Tip-tracking of EB1-GFP on the same dynamic microtubule as (A) visualized by TIRF microscopy.

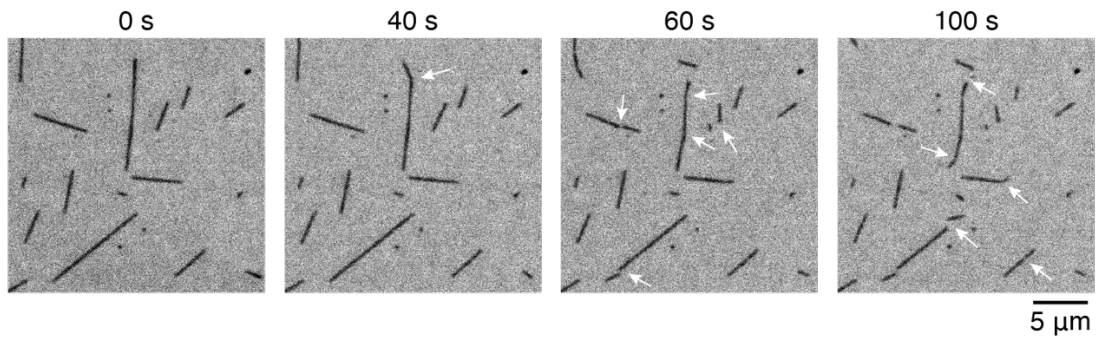

**Figure 3. Microtubule severing assay using IRM.** Time series of GMPCPP-stabilized microtubules severed by 3.5 nM of spastin imaged by IRM as described in [48]. Severing events can be detected from the breakages, kinks or gaps of the microtubule filaments (white arrows).